\documentclass[aps,pra,twocolumn,superscriptaddress]{revtex4-2}
\usepackage{color}
\usepackage{array}
\usepackage{textcomp}
\usepackage{bm}
\usepackage{braket}
\usepackage{amsmath}
\usepackage{amssymb}
\usepackage{upgreek}
\usepackage{multirow}
\usepackage{graphicx}
\usepackage{hyperref}
\hypersetup{colorlinks=true, citecolor=blue, urlcolor=blue, linkcolor=blue}

\begin{document}
\title{Enantioselective Orientation of Chiral Molecules Induced by Terahertz
Pulses with Twisted Polarization}
\author{Ilia Tutunnikov}
\thanks{These authors contributed equally to this work.}
\affiliation{AMOS and Department of Chemical and Biological Physics, The Weizmann Institute of Science, Rehovot 7610001, Israel}
\author{Long Xu}
\thanks{These authors contributed equally to this work.}
\affiliation{AMOS and Department of Chemical and Biological Physics, The Weizmann Institute of Science, Rehovot 7610001, Israel}
\author{Robert W. Field}
\email{rwfield@mit.edu}
\affiliation{Department of Chemistry, Massachusetts Institute of Technology, Cambridge, MA 02139,USA}
\author{Keith A. Nelson}
\email{kanelson@mit.edu}
\affiliation{Department of Chemistry, Massachusetts Institute of Technology, Cambridge, MA 02139,USA}
\author{Yehiam Prior}
\email{yehiam.prior@weizmann.ac.il}
\affiliation{AMOS and Department of Chemical and Biological Physics, The Weizmann Institute of Science, Rehovot 7610001, Israel}
\author{Ilya Sh. Averbukh}
\email{ilya.averbukh@weizmann.ac.il}
\affiliation{AMOS and Department of Chemical and Biological Physics, The Weizmann Institute of Science, Rehovot 7610001, Israel}

\begin{abstract}
Chirality and chiral molecules are key elements in modern chemical and biochemical industries. Individual addressing, and the eventual separation of chiral enantiomers has been and still is an important elusive task in molecular physics and chemistry, and a variety of methods has been introduced over the years to achieve this goal. Here, we theoretically demonstrate that a pair of cross-polarized THz pulses interacting with chiral molecules through their permanent dipole moments induces an enantioselective orientation of these molecules. This orientation persists for a long time, exceeding the duration of the THz pulses by several orders of magnitude, and its dependency on temperature and pulses' parameters
is investigated. The persistent orientation may enhance the deflection of the molecules in inhomogeneous electromagnetic fields, potentially leading to viable separation techniques.
\end{abstract}
\maketitle

\section{Introduction \label{sec:Introduction}}
Molecular chirality was discovered in the 19th century by Louis Pasteur \citep{Pasteur1848} (for a historical excursion, see e.g. \citep{Mason2007}),
and since then chirality has attracted considerable interest owing
to its importance in physics, chemistry, biology, and medicine. Chiral
molecules exist in two forms, called left-, and right-handed enantiomers,
which are mirror images of each other and cannot be superimposed
by translations or rotations \citep{Cotton1990Chemical}. Even
though the two enantiomers have many identical chemical and physical
properties, e.g. boiling points and moments of inertia, they often
differ in their biological activities. This is only one of the reasons
why the abilities to differentiate, selectively manipulate, and separate
the enantiomers in mixtures containing both of them are of great practical importance. From the point of view of fundamental physics, chiral molecules seem to be promising candidates for experiments aimed at measuring parity violation effects \citep{Berger2019}.

Some of the physical properties of the two enantiomers mimic their reflection relation. For example,
the product of three Cartesian components of the molecular dipole moment has the opposite
sign for the two enantiomers \citep{patterson2013enantiomer}. Utilizing the distinct properties of the enantiomers, a variety
of methods for chiral discrimination using electromagnetic fields
has been developed over the years, including: microwave three-wave
mixing spectroscopy \citep{patterson2013enantiomer,Patterson2013Sensitive,Patterson2014New,Alvin2014Enantiomer,lehmann2018theory,leibscher2019principles},
photoelectron circular dichroism \citep{Ritchie1976Theory,Bowering2001Asymmetry,Lux2011Circular,Beaulieu2017Attosecond,Beaulieu2018Photoexcitation},
Coulomb explosion imaging \citep{Pitzer2013Direct,Herwig2013Imaging,Fehre2019Enantioselective},
and high-order harmonic generation \citep{Cireasa2015Probing}. Most
recently, a generic purely optical method for enantioselective \emph{orientation}
has been proposed \citep{Yachmenev2016Detecting,Gershnabel2018Orienting,Tutunnikov2018Selective,Tutunnikov2019Laser}
and experimentally demonstrated on propylene oxide molecule spun by
an optical centrifuge \citep{Milner2019Controlled} in which the polarization vector rotates unidirectionally with increasing angular velocity.
This last method utilizes laser fields with ``twisted'' polarization [see Fig. \ref{fig:THz-field}(a)] and relies
on the off-diagonal elements of the polarizability tensor which, in
chiral molecules, have opposite signs for the two enantiomers.
In related works, it was demonstrated that the optically induced orientation
persists long after the end of the laser pulses \citep{Tutunnikov2019Laser,Tutunnikov2020Observation}.
For the purpose of the current work, fields with twisted polarization are understood to include: pairs of delayed
cross-polarized pulses \citep{Fleischer2009Controlling,Kitano2009Ultrafast,Khodorkovsky2011} [see Fig. \ref{fig:THz-field}(b)],
chiral pulse trains \citep{Zhdanovich2011Control,Johannes2012Molecular},
polarization-shaped pulses \citep{Kida2008,Kida2009,Karras2015Polarization,Prost2017Third,Mizuse2020},
and the aforementioned optical centrifuge for molecules \citep{Karczmarek1999Optical,Villeneuve2000Forced,Yuan2011Dynamics,Korobenko2014Direct,Korobenko2018Control}.

In this work, we study the orientation of chiral molecules induced
by terahertz (THz) pulses with twisted polarization. Since the advent
of THz pulse technology, intense THz pulses have been exploited for producing transient
field-free orientation of polar molecules of various complexity \citep{Fleischer2011Molecular,Kitano2013Orientation,Egodapitiya2014Terahertz,Babilotte2016Observation,Coudert2017Optimal,Damari2017Coherent,Tehini2019Shaping}.
Although transient orientation revival spikes may periodically appear on long time scales,
the orientation signature generally rides on a zero baseline and its time average is exactly
zero. Recently, it was shown that a single linearly polarized THz
pulse induces persistent orientation in symmetric- and asymmetric-top
molecules, including chiral molecules \citep{xu2020longlasting}.
In contrast to the transient signal, persistent orientation means that the time-averaged
post-pulse orientation degree differs from zero on a  long time scale, in the case discussed in this paper - exceeding the duration of the THz pulse by several orders of magnitude.

Here, we theoretically demonstrate that when applied to \emph{chiral molecules}, a \emph{pair of time-delayed cross-polarized THz pulses} induces orientation
in a direction perpendicular to the plane spanned by the polarizations.
The orientation is enantioselective, meaning that the two enantiomers
are oriented in opposite directions relative to the plane. We show that
the time-averaged projections of the molecular dipole moment onto all
three laboratory axes remains nonzero on a nanosecond timescale.
The paper is organized as follows. In Sec. \ref{sec:Methods}, we
briefly summarize our theoretical methods. In Sec. \ref{sec:Results},
we present and analyze the results of classical and fully quantum simulations
of THz field-driven molecular rotational dynamics, and Section \ref{sec:conclusions}
concludes the presentation.

\section{Methods \label{sec:Methods}}

We carried out classical as well as fully quantum mechanical simulations
of the THz field-driven rotational dynamics of chiral molecules. This
section outlines the theoretical approaches used in both cases.\\

\noindent \textbf{Quantum simulation.} The Hamiltonian describing
the molecular rotation driven by a THz field interacting with the
molecular dipole moment is given by \citep{Krems2018Molecules,Koch2019Quantum}
\begin{equation}
\hat{H}(t)=\hat{H}_{r}+\hat{H}_{\mathrm{int}}(t),\label{eq:Rotational-Hamiltonian}
\end{equation}
where $\hat{H}_{r}$ is the rotational kinetic energy Hamiltonian
and $\hat{H}_{\mathrm{int}}(t)=-\hat{\bm{\mu}}\cdot\mathbf{E}(t)$
is the molecule-field interaction. Here $\hat{\bm{\mu}}$ is the molecular
dipole moment operator and $\mathbf{E}(t)$ is the external electric
field. In this work, the contributions of higher order interaction
terms are small, and are not included. For the quantum mechanical treatment, it is
convenient to express the Hamiltonian in the basis of field-free symmetric-top wave functions $|JKM\rangle$ \citep{zare1988Angular}. Here $J$
is the value of the total angular momentum (in units of $\hbar$),
while $K$ and $M$ are the values of the projections on the molecule-fixed
axis (here the axis with smallest moment of inertia) and the laboratory-fixed
$Z$ axis, respectively. The nonzero matrix elements of the asymmetric-top
kinetic energy operator are given by \citep{zare1988Angular}{\small{}
\begin{flalign}
\langle JKM|\hat{H}_{r}|JKM\rangle=\frac{B+C}{2} & \left[J(J+1)-K^{2}\right]+AK^{2},\label{eq:HR1}\\
\langle JKM|\hat{H}_{r}|J,K\pm2,M\rangle & =\frac{B-C}{4}f(J,K\pm1),\label{eq:HR2}
\end{flalign}
}where $f(J,K)=\sqrt{(J^{2}-K^{2})[(J+1)^{2}-K^{2}]},$ $A=\hbar^{2}/2I_{a},\,B=\hbar^{2}/2I_{b},\,C=\hbar^{2}/2I_{c}$
are the rotational constants ($A>B>C$), and $I_{a}<I_{b}<I_{c}$
are the molecular moments of inertia. The time-dependent Schr\"{o}dinger
equation $i\hbar\partial_{t}|\Psi(t)\rangle=\hat{H}(t)|\Psi(t)\rangle$
is solved by numerical exponentiation of the Hamiltonian matrix (see
Expokit \citep{sidje1998Expokit}), and a detailed description of our numerical scheme can be found in \citep{xu2020longlasting}. In our simulations,
the computational basis included all the rotational states with an
angular momentum $J\leq44$. For the case of propylene oxide molecules which is used as an example, at a temperature of $T=5\, \mathrm{K}$, initial states with $J\leq8$ are included in the thermal averaging.\\

\noindent \textbf{Classical simulation.} In the classical limit, chiral
molecules are modeled as asymmetric tops driven by an external torque.
The classical equations of motion for the angular velocities (written
in the frame of principal axes of inertia tensor) are the Euler's
equations \citep{Goldstein2002Classical}
\begin{equation}
\mathbf{I}\bm{\dot{\Omega}}=(\mathbf{I}\bm{\Omega})\times\bm{\Omega}+\mathbf{T},\label{eq:Eulers-equations}
\end{equation}
where $\bm{\Omega}=(\Omega_{a},\Omega_{b},\Omega_{c})$ is the angular
velocity vector, $\mathbf{I}=\mathrm{diag}(I_{a},I_{b},I_{c})$ is
the moment of inertia tensor, and $\mathbf{T}=(T_{a},T_{b},T_{c})$
is the external torque vector. The external torque originates from
the electric field, which is defined in the laboratory frame.
In order to solve Eq. (\ref{eq:Eulers-equations}), a time-dependent
relation between the molecular frame (the frame of principal axes of inertia) and the laboratory frame is required.

Such a relation can be established with the help of a time-dependent
unit quaternion, which is used to parametrize the rotation of the
rigid body. Quaternions extend complex numbers, and are defined as quadruples of real numbers,
$q=(q_{0},q_{1},q_{2},q_{3})$. The relation between a vector $\mathbf{x}$
in the molecular frame and a vector $\mathbf{X}$ in the laboratory
frame is given by a simple linear transformation $\mathbf{x}=Q(t)\mathbf{X},$
where $Q(t)$ is a $3\times3$ matrix composed of the quaternion's
elements \citep{Coutsias2004The,Kuipers1999Quaternions}. The quaternion
obeys the following equation of motion
\begin{equation}
\dot{q}=\frac{1}{2}q\Omega,\label{eq:Quaternion-equation-of-motion}
\end{equation}
where $\Omega=(0,\bm{\Omega})$ is a pure quaternion and the quaternion
multiplication rule is implied \citep{Coutsias2004The,Kuipers1999Quaternions}.
Equations (\ref{eq:Eulers-equations}) and (\ref{eq:Quaternion-equation-of-motion})
are coupled via the torque, $\mathbf{T}=\boldsymbol{\mu}\times Q\mathbf{E}$,
where $\boldsymbol{\mu}$ is the molecular dipole moment vector.

To simulate the behavior of a classical ensemble, we use the Monte
Carlo approach. For each individual asymmetric top, we numerically solve the system of Eqs.
(\ref{eq:Eulers-equations}) and (\ref{eq:Quaternion-equation-of-motion}).
We use ensembles consisting of $N=10^{7}$ molecules, which are initially
isotropically distributed in space, and the initial angular velocities
are given by the Boltzmann distribution
\begin{equation}
P(\bm{\Omega})\propto\exp\left(-\frac{\bm{\Omega}^{T}\mathbf{I}\bm{\Omega}}{2k_{B}T}\right)=\prod_{i}\exp\left(-\frac{I_{i}\Omega_{i}^{2}}{2k_{B}T}\right),\label{eq:Classical-Boltzmann-distribution}
\end{equation}
where $T$ is the temperature and $k_{B}$ is the Boltzmann constant.
The initial uniform random quaternions were generated using the recipe
described in \citep{Lavalle2006Planning}.

\section{Results \label{sec:Results}}

\begin{figure}[t]
\centering{}\includegraphics[width=\columnwidth]{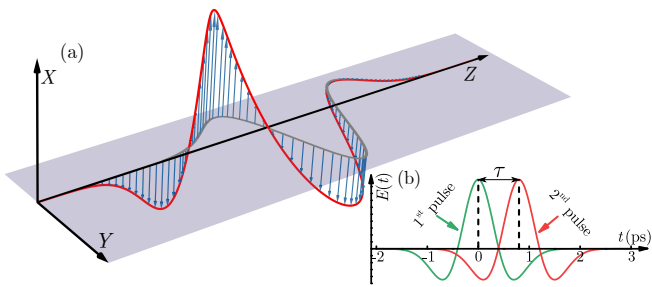}
\caption{(a) Illustration of the electric field with twisted polarization composed
of two delayed cross-polarized THz pulses {[}see Eq. (\ref{eq:THz-field}){]}.
The vertical blue arrows represent the $X$-projection of the field.
The gray line in the $YZ$ plane represents the $Y$-projection of
the field. (b) Amplitude of the $X$-polarized (green) and the $Y$-polarized (red)
pulses. Time dependence of the first pulse is given by $f(t)=(1-2at^{2})e^{-at^{2}}$,
while that of the second pulse by $f(t-\tau)$. Here, $a=3.06\,\mathrm{ps^{-2}}$
and $\tau=0.80\,\mathrm{ps}$. Time integral of the electric field
is identically zero. \label{fig:THz-field}}
\end{figure}

\begin{table}[!b]
\begin{centering}
\begin{tabular}{>{\centering}m{1.8cm}|>{\centering}m{2cm}|>{\centering}m{2.5cm}}
{\scriptsize{}Molecule\\} & {\scriptsize{}Moments of inertia} & {\scriptsize{}Molecular dipole components}\tabularnewline
\hline
\multirow{3}{1.8cm}{{\scriptsize{}(\emph{R})-propylene oxide}} & {\scriptsize{}$I_{a}=180386$} & {\scriptsize{}$\mu_{a}=0.965$}\tabularnewline
 & {\scriptsize{}$I_{b}=493185$} & {\scriptsize{}$\mu_{b}=-1.733$}\tabularnewline
 & {\scriptsize{}$I_{c}=553513$} & {\scriptsize{}$\mu_{c}=0.489$}\tabularnewline
\end{tabular}
\par\end{centering}
\caption{Molecular properties of (\emph{R})-PPO: eigenvalues of the moment
of inertia tensor (in atomic units) and components of dipole moment
(in Debye) in the frame of molecular principal axes of inertia. \label{tab:Molecular-properties}}
\end{table}

We consider propylene oxide (referred to as PPO hereafter) as a typical example of a chiral
molecule. Table \ref{tab:Molecular-properties} summarizes the molecular
properties of the right-handed enantiomer, (\emph{R})-PPO. Molecular
moments of inertia and the components of molecular dipole moments
were computed with the help of the GAUSSIAN software package (CAM-B3LYP/aug-cc-pVTZ
method) \citep{Frisch2016Gaussian}.

The molecules are excited by a pair of delayed cross-polarized THz pulses. The combined electric
field of the pulses is modeled using \citep{Coudert2017Optimal}
\begin{equation}
\mathbf{E}(t)=E_{0}\left[f(t)\mathbf{e}_{X}+f(t-\tau)\mathbf{e}_{Y}\right],\label{eq:THz-field}
\end{equation}
where $E_{0}$ is the peak amplitude, $f(t)=(1-2at^{2})e^{-at^{2}}$
defines the time dependence of each pulse, $a$ determines the temporal width of the pulse, $\tau$ is the time delay between the peaks of the two
pulses, and $\mathbf{e}_{X}$ and $\mathbf{e}_{Y}$ are the unit vectors
along the laboratory $X$ and $Y$ axes, respectively. The pulses
propagate along the laboratory $Z$ axis, while $\mathbf{E}$ twists
in the $XY$ plane, as shown in Fig. \ref{fig:THz-field}.

\begin{figure}
\centering{}\includegraphics[width=0.95\linewidth]{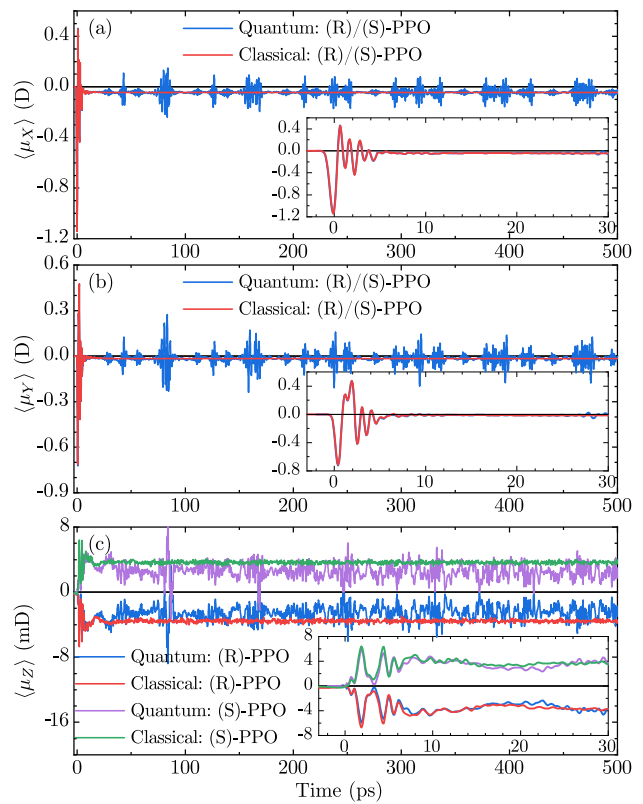}
\caption{Projections of the molecular dipole moment on the laboratory-fixed
axes. The classical results (red and green) are ensemble averages,
while quantum results (blue and purple) are incoherent averages over
the initial thermally populated rotational states. Here $T=5\,\mathrm{K}$,
and the THz field parameters are similar to Fig. \ref{fig:THz-field},
with $E_{0}=8.0\,\mathrm{MV/cm}$. The insets show the magnified portions
of the plots during the first 30 ps. \label{fig:Quantum-classical-0.5-ns}}
\end{figure}

\begin{figure}[t]
\centering{}\includegraphics[width=0.95\linewidth]{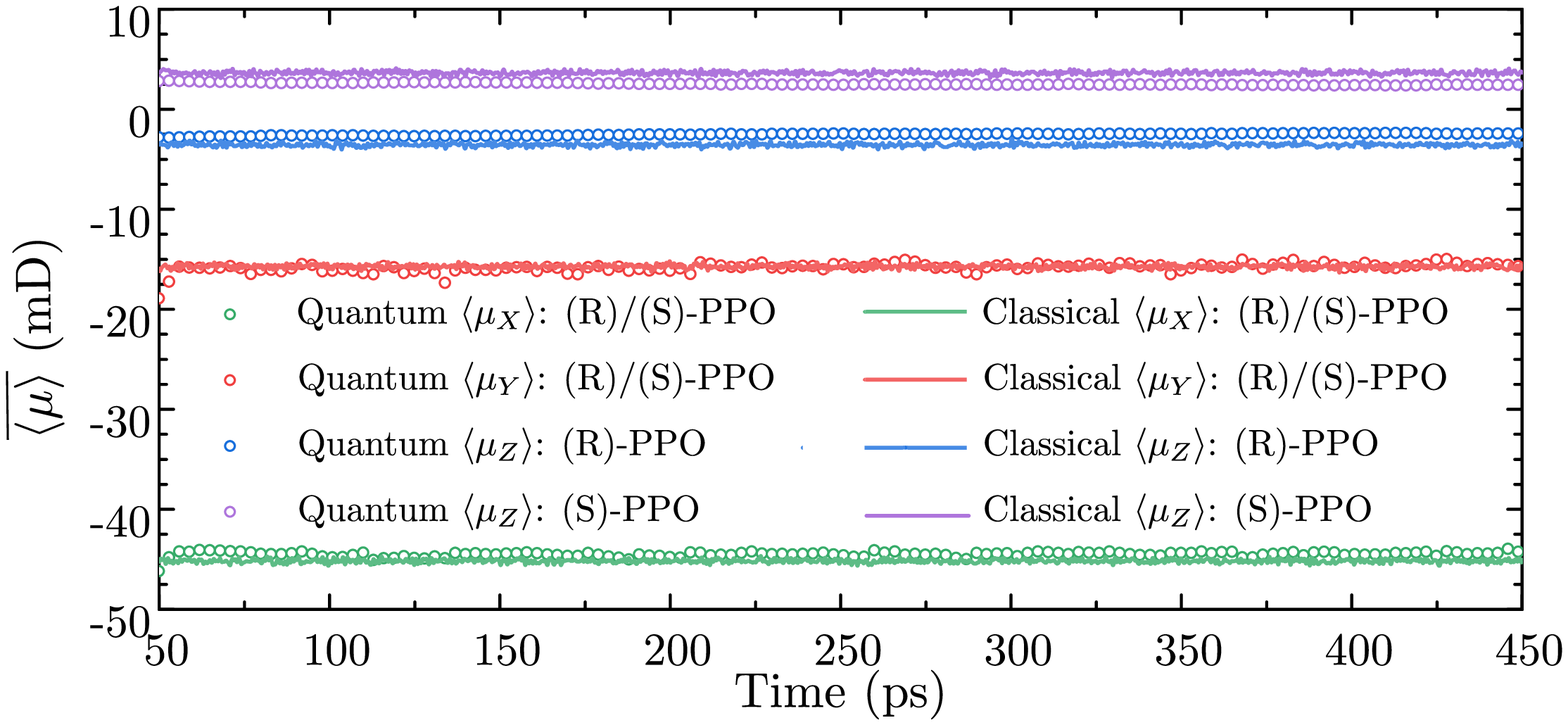}
\caption{Long-term behavior of the dipole projections shown in Fig. \ref{fig:Quantum-classical-0.5-ns}.
Open circles represent sliding time average of the dipole signals
$\braket{\mu_{X}}$ (green), $\braket{\mu_{Y}}$ (red), and $\braket{\mu_{Z}}$
(blue and purple) calculated quantum mechanically. The sliding time
average is evaluated according to $\overline{\langle\mu_{i}\rangle(t)}=(\Delta t)^{-1}\int_{t-\Delta t/2}^{t+\Delta t/2}\mathrm{d}t'\langle\mu_{i}\rangle(t')$,
where $\Delta t=100$ ps. For comparison, the correspondingly colored
solid lines are the classical results. \label{fig:Averaged-Quantum-classical}}
\end{figure}

\begin{figure*}
\begin{centering}
\includegraphics[width=0.95\linewidth]{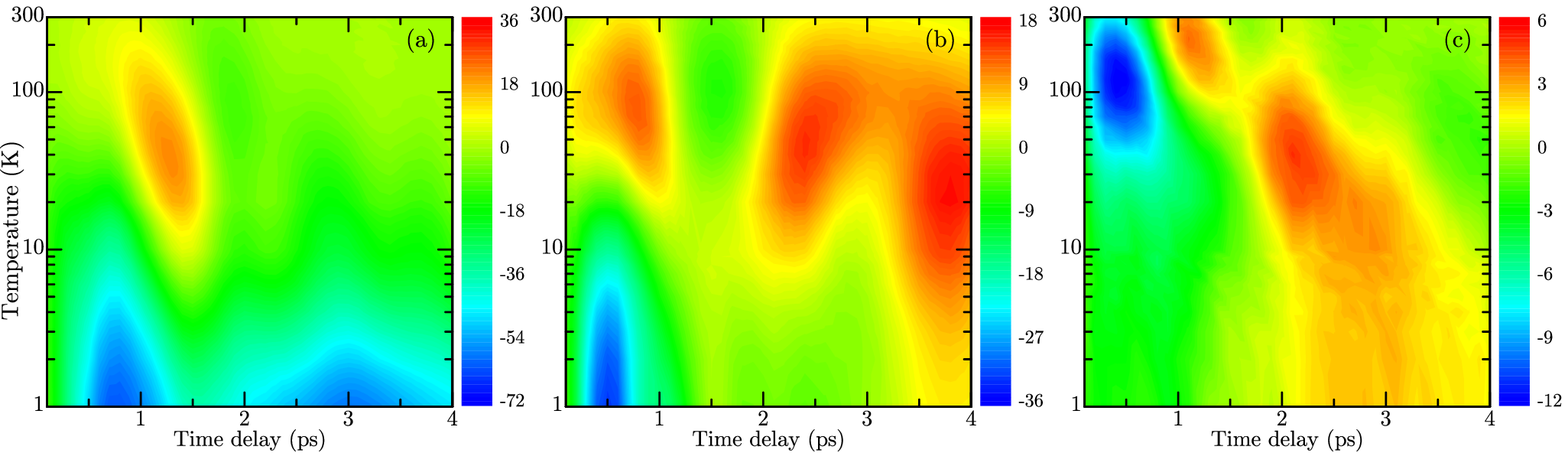}
\par\end{centering}
\caption{Classically calculated permanent values of the dipole signals
(a) $\langle\mu_{X}\rangle_{p}$, (b) $\langle\mu_{Y}\rangle_{p}$,
and (c) $\langle\mu_{Z}\rangle_{p}$ as functions of temperature,
$T$, and time delay between the two THz pulses, $\tau$. The color
scales are in the units of millidebye (mD). \label{fig:Long-time-dipole-delay-vs-temp}}
\end{figure*}

Figure \ref{fig:Quantum-classical-0.5-ns} shows the dipole moment
projections along the three laboratory axes, $\braket{\mu_{X}}$,
$\braket{\mu_{Y}}$, and $\braket{\mu_{Z}}$, as functions of time.
The angle brackets $\braket{\cdot}$ denote the incoherent average
of initial thermally populated rotational states, or the ensemble
average in the classical case. The parameters of the field are $a=3.06\,\mathrm{ps^{-2}}$,
$\tau=0.80\,\mathrm{ps}$, and $E_{0}=8.0\,\mathrm{MV/cm}$ (corresponding
to the peak intensity of $8.5\times10^{10}\,\mathrm{W/cm^{2}}$),
see Eq. (\ref{eq:THz-field}) and Fig. \ref{fig:THz-field}. Note that
THz pulses with peak amplitudes of tens of $\mathrm{MV/cm}$, especially with the use of field enhancement structures \citep{Clerici2013Wavelength,Oh2014Generation,Kuk2016Generation}, are experimentally available.
It is evident from the insets in Fig. \ref{fig:Quantum-classical-0.5-ns}
that on the short time scale the classical and quantum results are
in excellent agreement. Each of the THz pulses is followed by a splash
of dipole signal in the direction of the pulse polarization, i.e. initially along
the $X$ axis, and then along the $Y$ axis [see the minima in the insets of Figs. \ref{fig:Quantum-classical-0.5-ns}(a) and \ref{fig:Quantum-classical-0.5-ns}(b), which are before and after 0 ps, respectively]. This transient
orientation induced by single THz pulses is expected and has been observed before \citep{Fleischer2011Molecular,Kitano2013Orientation,Egodapitiya2014Terahertz,Babilotte2016Observation,Coudert2017Optimal,Damari2017Coherent,Tehini2019Shaping,xu2020longlasting}.

However, an unexpected result emerges: a dipole projection along the $Z$ axis (perpendicular to the plane of polarization twisting) appears after the second pulse {[}see Fig. \ref{fig:Quantum-classical-0.5-ns}(c){]}. This orientation which is unique to chiral molecules is enantioselective,
in the sense that the sign of the projection is opposite for the two enantiomers,
positive for (\emph{S})-PPO and negative for (\emph{R})-PPO. The enantioselectivity
of the orientation in the $Z$ direction can be formally derived as well {[}for details,
see Appendix \ref{sec:Appendix-Opposite-Sign}{]}. Similar enantioselective
orientation was observed in chiral molecules optically excited by laser fields with twisted polarization acting on the molecular polarizability \citep{Gershnabel2018Orienting,Tutunnikov2018Selective,Milner2019Controlled}.

Furthermore, the classical results clearly show another remarkable
feature of the induced orientation. After the field is switched off {[}$t>2.5$ ps, see Fig.
\ref{fig:THz-field}(b){]}, the dipole projections along all three
laboratory axes persist on the nanosecond timescale. The direct quantum simulation deviates from the classical one on the long time
scale,
exhibiting quantum beats/revivals \citep{Averbukh1989,Felker1992Rotational,Robinett2004}.
Nevertheless, as can be seen from Fig. \ref{fig:Averaged-Quantum-classical},
on a timescale of 0.5 ns the time-averaged quantum signals reproduce
well the steady state dipole signals obtained by the classical calculation.

Classically, the persistent long-term orientation shown in Figs. \ref{fig:Quantum-classical-0.5-ns}
and \ref{fig:Averaged-Quantum-classical} is in fact permanent. In
the absence of external torques (external fields), in the laboratory
frame the angular momentum vector is conserved, while in the molecular
frame the angular momentum follows a fixed trajectory, which can be visualized using the
Binet construction \citep{Goldstein2002Classical}. Although the absolute
orientation of an asymmetric top in the laboratory frame never recurs, depending
on the energy and magnitude of the angular momentum, the projections
of the molecular principal axes $a$ or $c$ on the conserved angular
momentum vector have a constant sign \citep{Goldstein2002Classical}.
As a result, the attained asymptotic values of the orientation factors
do not change after the initial dephasing of the ensemble which, according
to Fig. \ref{fig:Quantum-classical-0.5-ns}, takes about 30 ps. On
the other hand, quantum mechanically, the notion of well-defined trajectories
is invalid, which
means that the orientation is simply \emph{long-lived} and eventually will change its sign. Since the kinetic energy Hamiltonian
{[}see Eq. (\ref{eq:HR2}){]} couples rotational states with different
$K$ quantum number, the quantum-mechanical asymmetric top does not
have permanently oriented eigenstates. Any initially oriented state
will eventually oscillate between being oriented and anti-oriented,
an effect known as dynamical tunneling \citep{Keshavamurthy2011Dynamical}.
Formally, for a quantum mechanical chiral rotor, one would expect no permanent orientation after the turn-off
of all external fields. However,
as we show here and as was shown in \citep{Tutunnikov2019Laser,Tutunnikov2020Observation},
the tunneling timescale may exceed the excitation timescale by orders
of magnitude.

Notice, the persistent orientation appearing in the directions of
each of the pulses (along the $X$ and $Y$ axes) does not rely on chirality.
It was recently shown that single THz pulses applied to symmetric-
and asymmetric-top molecules also induce persistent orientation
\citep{xu2020longlasting}. In contrast, both the appearance and permanency of the orientation along the propagation
direction which is perpendicular to both the $X$ and $Y$ axes depend on the chirality of the molecule. Specifically, these
effects rely on the lack of molecular symmetry, i.e. all three molecular dipole moment components must differ from zero, $\mu_{a},\,\mu_{b},\,\mu_{c}\neq0$
{[}see Appendix \ref{sec:Appendix-Orientation}{]}. For comparison,
in the case of optical excitation by the laser fields with twisted polarization,
the orientation relies on the existence of the off-diagonal elements
of the polarizability tensor, which is a property of chiral molecules
as well. In that case, the \emph{induced dipole} has nonzero projections
along all three principal axes, even when the laser field is
polarized along only one of the molecular principal axes.

The magnitude of the THz-induced orientation is sensitive to the initial
temperature and external field parameters \citep{Lapert2012Field,Shu2013Field,Tehini2019Shaping,xu2020longlasting}.
Through our classical simulations, we carried out an extensive
study of the permanent orientation dependence on temperature, $T$, and the time delay between the two cross-polarized pulses, $\tau$
{[}see Eq. (\ref{eq:THz-field}){]}. The results are summarized in
Fig. \ref{fig:Long-time-dipole-delay-vs-temp}. The permanent values
of the dipole projections (denoted by $\langle\mu_{X}\rangle_{p}$,
$\langle\mu_{Y}\rangle_{p}$, and $\langle\mu_{Z}\rangle_{p}$) were obtained
by following the field-free dynamics for a sufficiently long time
until a steady-state is reached (typically $t>100$ ps). Figure
\ref{fig:Long-time-dipole-delay-vs-temp} shows that for a given rotational
temperature, there are one or several disjoint ranges of $\tau$ resulting
in optimal (largest absolute value) orientation. In general, the optimal time
delay between the pulses is shorter for higher temperatures. Note, however, that the temperature dependence of the individual projections $\langle\mu_{X}\rangle_{p}$, $\langle\mu_{Y}\rangle_{p}$, and $\langle\mu_{Z}\rangle_{p}$ is non-monotonic. For example, at a fixed time delay $\tau\approx0.5\,\mathrm{ps}$,
$|\langle\mu_{Z}\rangle_{p}|$ increases with temperature up to $T\approx120\,\mathrm{K}$,
after which $|\langle\mu_{Z}\rangle_{p}|$ begins to decrease.

\section{conclusions \label{sec:conclusions}}

We theoretically demonstrated a qualitatively new phenomenon of field-free
enantioselective orientation of chiral molecules induced by THz pulses
with twisted polarization. The twisted pulse induces orientations
along the polarization directions of the two pulses forming the twisted
pulse, and this orientation is of the same sign for both enantiomers. In the direction perpendicular to the polarization
direction of both pulses, we find that the orientation is of opposite sign for the two enantiomers.
The latter effect relies on the molecular chirality, namely on the lack of molecular symmetry,
such that the molecular dipole has nonzero projections on all three molecular principal axes.
The orientation was shown to persist long after the end of the pulses.
We studied the dependence of persistent orientation values on the
temperature and the time delay between the two cross-polarized THz pulses.
The orientation factors were found to be quite robust against the detrimental
effects of temperature provided that the time delay is adjusted appropriately.
The orientation dynamics on timescales beyond nanoseconds requires
a more detailed analysis, as other effects such as collisions and fine structure
effects \citep{Thomas2018Hyperfine} become important in addition to dynamical tunneling. The relative importance of such
effects should be assessed in future works. The orientation persisting
on the nanosecond timescale may be measured by means of second (or
higher order) harmonic generation \citep{Frumker2012Oriented}, and
could be used for deflection by inhomogeneous electric fields \citep{Gershnabel2011Electric,Gershnabel2011Deflection}
(for an extensive review, see \citep{Chang2015}). The enantioselective
orientation along the propagation direction may be useful for fast and precise analysis of enentiomeric excess, and may facilitate \emph{enantioselective} separation using
inhomogeneous fields \citep{Yachmenev2019Field}. In the past \citep{Yachmenev2016Detecting, Gershnabel2018Orienting, Tutunnikov2018Selective, Tutunnikov2019Laser, Milner2019Controlled, Tutunnikov2020Observation}, related effects induced by optical pulses with twisted polarizations have been reported, and further exploration will examine the combined effect of THz and optical fields together that could maximize the difference in orientations between the two enantiomers.

\section*{Acknowledgments}
\noindent
I.A. gratefully acknowledges support by the Israel Science Foundation (Grant No. 746/15). K.A.N. acknowledges support by the U.S. National Science Foundation Grant CHE-1665383.
I.A. acknowledges support as the Patricia Elman Bildner Professorial Chair. This research was made possible in part by the historic generosity of the Harold Perlman Family.

\appendix

\section{Qualitative demonstration of enantioselective orientation \label{sec:Appendix-Opposite-Sign}}

\begin{figure*}
\begin{centering}
\includegraphics[width=8cm]{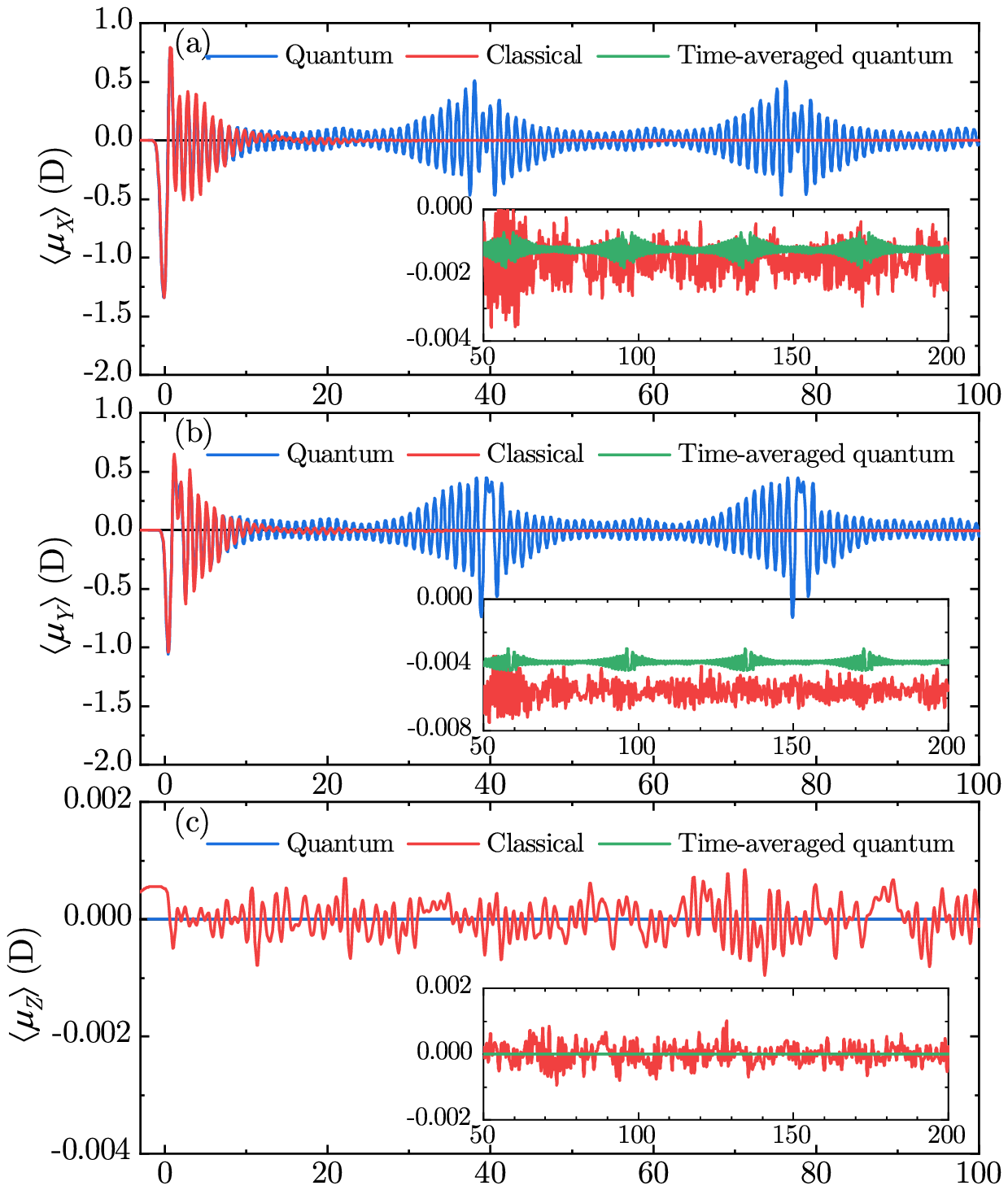} \includegraphics[width=8cm]{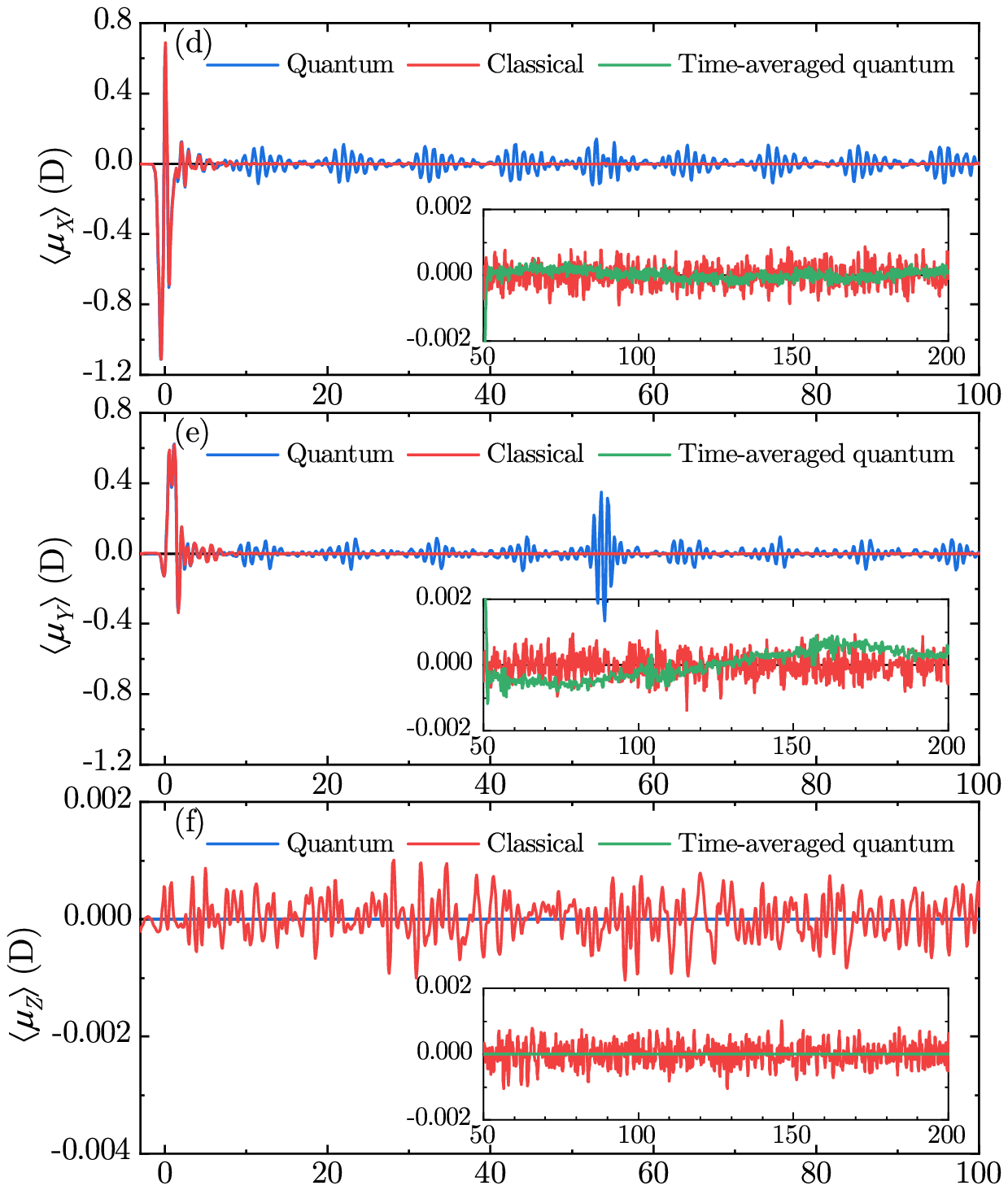}
\par\end{centering}
\caption{Thermally averaged $X,Y,Z$-projections of the dipole moment in the
laboratory frame, $\braket{\mu_{X}}$, $\braket{\mu_{Y}}$, and $\braket{\mu_{Z}}$
as functions of time for (a-c) $\mathrm{CH_{3}Cl}$ and (d-f) $\mathrm{SO_{2}}$
molecules. The results of quantum and classical simulations are shown
in blue and red lines, respectively. The green curve represents the
sliding time average defined by $\overline{\langle\mu_{i}\rangle(t)}=(\Delta t)^{-1}\int_{t-\Delta t/2}^{t+\Delta t/2}\mathrm{d}t'\langle\mu_{i}\rangle(t')$,
where (a-c) $\Delta t=38.2$ ps and (d-f) $\Delta t=100$ ps, respectively.
The insets show a magnified portion of the signals. \label{fig:CH3Cl_SO2}}
\end{figure*}

Consider two overlapping THz pulses propagating along the laboratory $Z$ axis, and which are polarized along $X$ and $Y$ axes, respectively. The nonzero matrix
elements of the interaction potential can be written as
\begin{align}
 & \langle JKM|\hat{H}_{\mathrm{int}}(t)|J',K-s,M-p\rangle\nonumber \\
= & \mu_{s}^{(1)}E_{-p}^{(1)}(t)\langle JKM|{D_{p,s}^{1}}^{*}(R)|J',K-s,M-p\rangle,\label{eq:Hint2}
\end{align}
where $s=0,\pm1$, and $p=\pm1$ (since $E_{0}^{(1)}=0$). Here $\mu_{s}^{(1)}$
and $E_{p}^{(1)}$ are the dipole moment and electric field, respectively,
expressed as spherical tensors of rank 1, with components $\mu_{\pm1}^{(1)}=\mp(\mu_{b}\pm i\mu_{c})/\sqrt{2}$,
$\mu_{0}^{(1)}=\mu_{a}$, and $E_{\pm1}^{(1)}=\mp(E_{X}\pm iE_{Y})/\sqrt{2}$,
$E_{0}^{(1)}=E_{Z}$.
${D_{p,s}^{1}}^{*}(R)$
is the complex conjugate of the Wigner D-matrix, where $R$ denotes
the set of the three Euler angles, $(\theta,\phi,\chi)$.

The THz pulses induce dynamics of the $Z$-component of the polarization, defined by
\begin{equation}
P_{Z}(t)=\sum\limits _{s=-1}^{+1}\mu_{s}^{(1)}\langle\Psi(t)|{D_{0,s}^{1}}^{*}(R)|\Psi(t)\rangle,\label{eq:Z-Polarization}
\end{equation}
where the wave function is given by $|\Psi(t)\rangle=\hat{U}(t,0)|\Psi(0)\rangle,$
with $\hat{U}(t,0)$ being the evolution operator. Since the discussion here is qualitative, for simplicity we assume that initially the rotor is in the ground rotational
state $\ket{JKM}=\ket{000}$, such that $\ket{\Psi(t)}=\hat{U}(t,0)|000\rangle$.

The evolution operator, $\hat{U}(t,0)$ can be expanded in a Dyson
series
\begin{widetext}
\begin{equation}
\hat{U}(t,0)=\sum\limits _{n=0}^{\infty}\frac{(-i)^{n}}{n!}\int_{0}^{t}\mathrm{d}t_{1}\int_{0}^{t}\mathrm{d}t_{2}\cdots\int_{0}^{t}\mathrm{d}t_{n}\hat{\mathcal{T}}\hat{H}_{\mathrm{int}}(t_{1})\hat{H}_{\mathrm{int}}(t_{2})\cdots\hat{H}_{\mathrm{int}}(t_{n}),\label{EqRef: Propagator}
\end{equation}
where $\hat{\mathcal{T}}$ is the time-ordering operator. From the properties of Wigner 3-j symbols it follows that the only
non-zero matrix elements, $\langle JKM|{D_{0,s}^{1}}^{*}(R)|J',K-s,M\rangle\neq0$,
are those satisfying $|J-1|\leq J'\leq J+1$. Hence, the nonzero components
of polarization are
\begin{align}
V_{s} & \sim\mu_{s}^{(1)}\langle\prod\limits _{n}\hat{H}_{\mathrm{int}}(t_{n})000|{D_{0,s}^{1}}^{*}(R)|\prod\limits _{n'}\hat{H}_{\mathrm{int}}(t_{n'}')000\rangle\nonumber \\
 & =\mu_{s}^{(1)}\sum\limits _{J_{1}^{\prime}J_{1}K_{1}M_{1}}\langle\prod\limits _{n}\hat{H}_{\mathrm{int}}(t_{n})000|J_{1}K_{1}M_{1}\rangle\langle J_{1}K_{1}M_{1}|{D_{0,s}^{(1)*}}|J_{1}^{\prime},K_{1}-s,M_{1}\rangle\langle J_{1}^{\prime},K_{1}-s,M_{1}|\prod\limits _{n'}\hat{H}_{\mathrm{int}}(t_{n'}^{\prime})000\rangle.\label{eq:Component}
\end{align}
\end{widetext}
Note that for the qualitative argument here, the evolution operator
in Eq. (\ref{eq:Component}) is considered as $\hat{U}(t,0)\sim\prod\limits _{n}\hat{H}_{\mathrm{int}}(t_{n})$.
From Eq. (\ref{eq:Component}), we know that $V_{s}$ differs from zero
only if $\langle J_{1}+J_{1}^{\prime},2K_{1}-s,2M_{1}|\prod\limits _{n}\hat{H}_{\mathrm{int}}(t_{n})\prod\limits _{n'}\hat{H}_{\mathrm{int}}(t_{n'}')|000\rangle\neq0$.
Also, according to Eq. (\ref{eq:Hint2}), the allowed transitions
are between the states with $M$ and $M\pm1$, because $p=\pm1$.
Hence, it follows from Eq. (\ref{eq:Component}) that the state changing
from $0$ to $2M_{1}$ involves an even number of interaction terms
($\hat{H}_{\mathrm{int}}$). Similarly, transitions between states
with $K$ quantum number equals $0$ and $2K_{1}-s$ involve $2l+s$
interaction terms with $s=0$ {[}see Eq. (\ref{eq:Hint2}), with $s=0${]}
and $2L+s$ interaction terms with $s=\pm1$ {[}see Eq. (\ref{eq:Hint2}),
with $s=\pm1${]}, where $L$ and $l$ are integers. The dipole moments
of two enantiomers of the chiral molecule satisfy $\mu_{\pm1}^{(S)}=\mu_{\pm1}^{(R)}$
and $\mu_{0}^{(S)}=-\mu_{0}^{(R)}$, such that the non-zero polarization
components {[}see Eq. (\ref{eq:Component}){]} satisfy
\begin{align}
\frac{V_{s}^{(S)}}{V_{s}^{(R)}} & =\frac{\mu_{s}^{(S)}}{\mu_{s}^{(R)}}\left(\frac{\mu_{-1}^{(S)}}{\mu_{-1}^{(R)}}\right)^{L^{\prime}}\left(\frac{\mu_{0}^{(S)}}{\mu_{0}^{(R)}}\right)^{2l+s}\left(\frac{\mu_{1}^{(S)}}{\mu_{1}^{(R)}}\right)^{2L+s-L^{\prime}}\nonumber \\
 & =\frac{\mu_{s}^{(S)}}{\mu_{s}^{(R)}}\left(\frac{\mu_{0}^{(S)}}{\mu_{0}^{(R)}}\right)^{s}=-1,\label{eq:Ratio}
\end{align}
where $s=0,\pm1$, and $L^{\prime}$ is an integer. As one can see, the components
of polarization of the two enantiomers are of opposite signs, and
thereby the polarizations satisfy $P_{Z}^{(S)}(t)=-P_{Z}^{(R)}(t)$.\\

\section{Orientation of non-chiral molecules
\label{sec:Appendix-Orientation}}

\begin{table}[!b]
\begin{centering}
\begin{tabular}{>{\centering}m{2cm}|>{\centering}m{2cm}|>{\centering}m{2.5cm}}
{\scriptsize{}Molecule\\} & {\scriptsize{}Moments of inertia} & {\scriptsize{}Molecular dipole components}\tabularnewline
\hline
\multirow{3}{2.cm}{{\scriptsize{}methyl chloride}} & {\scriptsize{}{}{}{}$I_{a}=20910$} & {\scriptsize{}{}{}{}$\mu_{a}=1.986$}\tabularnewline
 & {\scriptsize{}{}{}{}$I_{b}=251506$} & {\scriptsize{}{}{}{}$\mu_{b}=0$}\tabularnewline
 & {\scriptsize{}{}{}{}$I_{c}=251506$} & {\scriptsize{}{}{}{}$\mu_{c}=0$}\tabularnewline
 \hline
\multirow{3}{1.8cm}{{\scriptsize{}sulfur dioxide}} & {\scriptsize{}$I_{a}=57812$} & {\scriptsize{}$\mu_{a}=0$}\tabularnewline
 & {\scriptsize{}$I_{b}=323141$} & {\scriptsize{}$\mu_{b}=-1.805$}\tabularnewline
 & {\scriptsize{}$I_{c}=380953$} & {\scriptsize{}$\mu_{c}=0$}\tabularnewline
\end{tabular}
\par\end{centering}
\caption{Molecular properties of methyl chloride ($\mathrm{CH_{3}Cl}$) and sulfur dioxide ($\mathrm{SO_{2}}$) molecules: eigenvalues of the moment
of inertia tensor (in atomic units) and components of dipole moment
(in Debye) in the frame of molecular principal axes of inertia. \label{tab:Molecular-properties-CH3Cl-SO2}}
\end{table}

Figure \ref{fig:CH3Cl_SO2} shows the orientation signals along the
three laboratory axes, $\braket{\mu_{X}}(t)$, $\braket{\mu_{Y}}(t)$,
and $\braket{\mu_{Z}}(t)$, for two non-chiral molecules. The first one
is the symmetric-top methyl chloride ($\mathrm{CH_{3}Cl}$), in which
the molecular dipole moment is along the symmetry axis, $a$ axis ($\mu_{b}=\mu_{c}=0$, see Table \ref{tab:Molecular-properties-CH3Cl-SO2}).
The second molecule, sulfur dioxide ($\mathrm{SO_{2}}$) is an asymmetric-top
molecule in which the molecular dipole moment is along the molecular
$b$ axis ($\mu_{a}=\mu_{c}=0$, see Table \ref{tab:Molecular-properties-CH3Cl-SO2}). The parameters of the THz pulses
are similar to Fig. \ref{fig:Quantum-classical-0.5-ns}. The initial
temperature is $T=5\,\mathrm{K}$. The quantum and classical results are in
good agreement. As expected, both the symmetric- {[}Figs. \ref{fig:CH3Cl_SO2}(a)
and \ref{fig:CH3Cl_SO2}(b){]} and non-chiral asymmetric-top {[}Figs. \ref{fig:CH3Cl_SO2}(d)
and \ref{fig:CH3Cl_SO2}(e){]} molecules immediately respond to the
$X$-polarized and $Y$-polarized
pulses. However, in contrast to chiral molecules {[}see Fig. \ref{fig:Quantum-classical-0.5-ns}{]},
the $Z$-projection of the dipole moment remains zero {[}see Figs.
\ref{fig:CH3Cl_SO2}(c) and \ref{fig:CH3Cl_SO2}(f){]}. Note that
$\braket{\mu_{Z}}$ calculated quantum mechanically is identically
zero, while the small amplitude oscillations appearing in the classical
results are due to the finite number ($N$) of molecules in the ensemble.
The lack of orientation in the cases of symmetric- or non-chiral asymmetric-top
molecules indicates that all three molecular dipole components ($\mu_{a},\,\mu_{b},\,\mu_{c}$) are
indeed required for inducing the perpendicular (in the $Z$ direction)
orientation. Other combinations were considered numerically as well
(not shown), e.g. asymmetric top molecule having two non-zero molecular
dipole components. In all cases, there is no perpendicular orientation,
i.e. $\braket{\mu_{Z}}=0$.

\end{document}